\documentclass[preprint]{vgtc}               

\graphicspath{{figures/}{pictures/}{images/}{./}} 

\usepackage{times}                     

\usepackage{tabu}                      
\usepackage{booktabs}                  
\usepackage{lipsum}                    
\usepackage{mwe}  
\usepackage{comment}   

\usepackage{mathptmx}                  

\onlineid{1132}

\vgtccategory{Research}

\vgtcinsertpkg

\title{\vspace*{2cm}Enhancing Job Interview Preparation Through Immersive Experiences Using Photorealistic, AI-powered Metahuman Avatars}

\author{Navid Ashrafi\thanks{e-mail: ashrafi@tu-berlin.de}\\ %
        \scriptsize Technical University of Berlin %
\and Francesco Vona\thanks{e-mail: francesco.vona@hshl.de}\\ %
     \scriptsize University of Applied Sciences Hamm-Lippstadt %
\and Carina Ringsdorf\thanks{e-mail: ringsdorf@campus.tu-berlin.de}\\ %
     \scriptsize Technical University of Berlin
\and Christian Hertel\thanks{e-mail: hertel@campus.tu-berlin.de}\\ %
     \scriptsize Technical University of Berlin
\and Luca Toni\thanks{e-mail: Lu.To12@campus.tu-berlin.de}\\ %
     \scriptsize Technical University of Berlin
\and Sarina Kailer\thanks{e-mail: s.kailer@campus.tu-berlin.de}\\ %
     \scriptsize Technical University of Berlin
\and Alice Bartels\thanks{e-mail: alice.bartels@campus.tu-berlin.de}\\ %
     \scriptsize Technical University of Berlin
\and Tanja Kojic\thanks{e-mail: tanja.kojic@tu-berlin.de}\\ %
     \scriptsize Technical University of Berlin
\and Jan-Niklas Voigt-Antons\thanks{e-mail: jan-niklas.voigt-antons@hshl.de}\\ %
     \scriptsize University of Applied Sciences Hamm-Lippstadt}

\teaser{
  \centering
  \includegraphics[width=\linewidth]{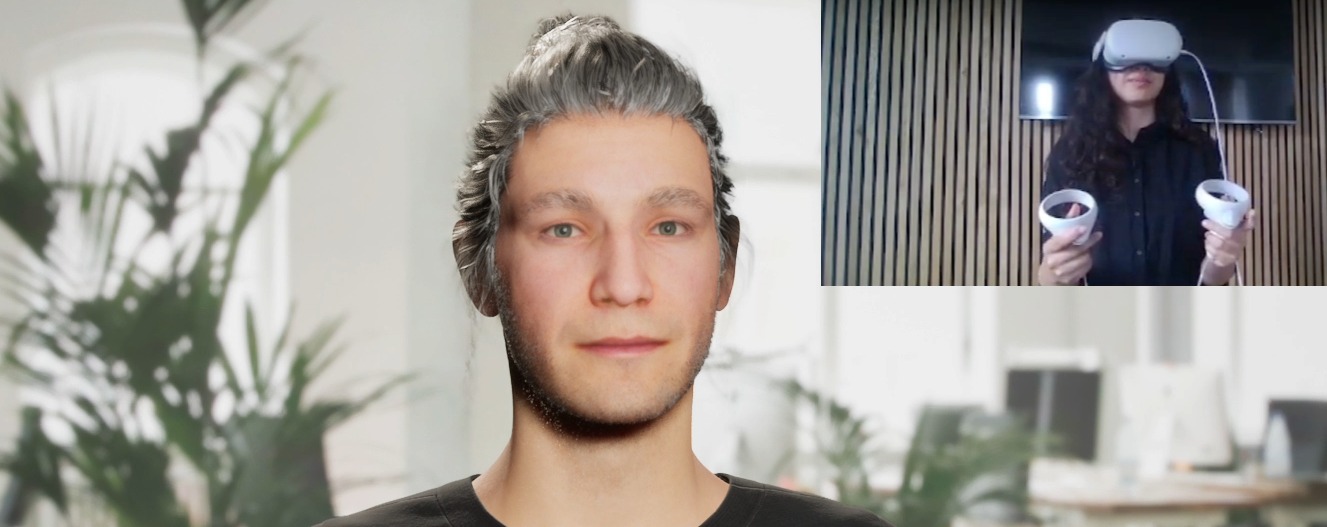}
  \caption{The VR setting of the software engineering job interview simulator, featuring \textit{"Mobert"}, the interviewer Metahuman.}
  \label{fig:teaser}
}

\abstract{
    This study will investigate the user experience while interacting with highly photorealistic virtual job interviewer avatars in Virtual Reality (VR), Augmented Reality (AR), and on a 2D screen. Having a precise speech recognition mechanism, our virtual character performs a mock-up software engineering job interview to adequately immerse the user in a life-like scenario. To evaluate the efficiency of our system, we measure factors such as the provoked level of anxiety, social presence, self-esteem, and intrinsic motivation. This research is a work in progress with a prospective within-subject user study including approximately 40 participants. All users will engage with three job interview conditions (VR, AR, and desktop) and provide their feedback. Additionally, Users' biophysical responses will be collected using a biosensor to measure the level of anxiety during the job interview.

} 


\begin{document}


\firstsection{Introduction}

\maketitle

The rapid growth of technology has led to increased competition in the software engineering job market and more strict acceptance criteria on soft and hard skills \cite{lulia}. Although self-training methods, e.g. rehearsing in front of a mirror 
could help one prepare for an upcoming job interview \cite{hard}, such methods may not build up a sufficient level of confidence and motivation, especially in students with little to no prior work experience \cite{lerner}. These individuals can greatly benefit from immersive experiences provoking a life-like level of anxiety such as a virtual mock-up job interview \cite{smith}.

Virtual reality (VR) applications have been shown to help people with certain types of social anxiety \cite{north}. Exposure therapy in VR has been extensively studied and has promised positive effects on users' ability to cope with stressful public speaking and presentations \cite{north}. These studies have revealed a potential correlation between avatars' realism and the sense of presence and anxiety. Related research on VR job interview simulations has also an improvement in users' anxiety control and social skills after the VR experience \cite{grill}. 
Stanica et al. \cite{lulia} introduced a VR job interview simulator featuring an interviewer avatar for university students. Although they have found their system to have a high level of modularization and accessibility, the study lacks a precise analysis of the user's state while using the system. In a similar research, Kwon et al. \cite{kwon} have found a correlation between the level of avatar photorealism and the social presence in their job interview application. 
While these studies have addressed certain aspects of job interviews in VR, the usage of Augmented reality (AR) has not been thoroughly investigated in this context. 
A few studies have investigated the use of AR in assisting individuals with autism and disabilities \cite{ar-au}. However, AR can also help students prepare for college or the job industry by improving their social, academic, and vocational skills \cite{kell}. Integrating highly photorealistic Metahuman avatars \cite{epic} in an AR setting and utilizing the benefits of AR, such as passthrough, our research aims to further enhance immersive job interview experiences for students. To the best of our knowledge, this study is the first to implement and compare VR, AR, and desktop versions of a software engineering job interview simulation featuring highly photorealistic Metahuman avatars 
We formulated the research question \textbf{"How would the level of provoked anxiety, sense of presence, self-esteem, and intrinsic motivation vary across different levels of reality(VR, AR, and desktop) in a mock-up software engineering job interview using photorealistic Metahumans"} and we accordingly hypothesized the following:
\begin{itemize}
    \item \textbf{H1:} The virtual settings (AR \& VR) would lead to a higher level of anxiety and presence compared to the desktop setting.
    \item \textbf{H2:} The users' provoked anxiety and sense of presence would be higher in the VR setting as compared to AR.
    \item \textbf{H3:} The virtual settings (AR \& VR) would lead to higher intrinsic motivation and self-esteem compared to the desktop condition with more significance in the AR setting than in VR. 
\end{itemize} 

\section{Mobert, the Interactive Metahuman}
We used \textit{Unreal Engine's} Metahuman creator \cite{epic} to design and sculpt our avatars. A substantial amount of research focuses on avatar embodiment and appearance \cite{looks}. Based on the existing guidelines, we designed the male Metahuman \textit{"Mobert"} with a friendly and competent appearance. We integrated our avatar with a Convai chatbot \cite{conv} to make it conversational. Although Convai is not open-source and has a high number of dependencies, their precise speech recognition (which works properly even in noisy environments) and pre-training functionalities, enable virtual characters across different platforms to perform human-like conversations. We implemented a push-to-talk button to record the user voice input, which will be transcribed and fed to the Convai chatbot. The response from the chatbot will be transmitted back to the avatar using a text-to-speech component. Utilizing Metahuman's lip-sync and face animations, the avatar can respond to the user with realistic lip movements and facial expressions. The application was initially built for a desktop computer and was then developed into a VR version compatible with the Meta Quest 3 headset (Figure \ref{fig:teaser}). An AR version of the same application is currently under development. Additionally, a female avatar is being designed for each condition to randomly replace the male interviewer to avoid gender bias. 

\section{Study design}

To verify our hypotheses, we designed a within-subject study where each participant takes part in all conditions (AR, VR, and desktop) in a randomized order. The VR and AR conditions run on a Meta Quest 3 headset, and the desktop version would display on a 4K screen. Each condition takes approximately 12 minutes in which the avatar welcomes the user to the interview and proceeds with asking 10 software engineering-related questions. The chatbot will adapt the follow-up questions to provide a natural conversational flow. Upon receiving irrelevant user input, the chatbot is trained to steer the conversation back to the purpose of the interview. To measure anxiety, we use \textit{Empatica Embrace Plus} wristbands while performing each condition which allows for recording electrodermal, cardiovascular, and skin temperature data. To assess the user's sense of presence, Lombard's Presence Inventory (TPI) \cite{ditton} is used that offer data collection on subscale categories of spatial, social, passive, and active presence. Information regarding self-esteem and intrinsic motivation will be collected via Rosenberg's Self-Esteem Scale (RSES) and the Intrinsic Motivation Inventory (IMI) \cite{imi}, including subscale categories such as perceived enjoyment and usefulness. The questionnaires will be filled out before the first and after each condition. Moreover, the Affinity for Technology Interaction (ATI) \cite{ati} and a demographics survey will be administered at the beginning of the study, and a questionnaire regarding user preferences among the three conditions will be used at the end of the experiment. 

Data derived from the mentioned questionnaires and the biosensor 
would provide sufficient evidence to examine our hypotheses. Based on the distribution of our datasets, we will use a suitable analysis method e.g. Repeated Measure ANOVA to reveal significance within our data. One challenging task is using the biosensor to accurately determine the level of anxiety as it is a complex emotion with a diverse spectrum and various influencing factors. Upon approval of our study by the university ethics committee, approximately 40 students with a basic technical background will be recruited and their participation will be compensated by 15 Euros.


\bibliographystyle{abbrv-doi}

\bibliography{template}

\begin{thebibliography}{10}

\bibitem{epic}
{Epic Games}.
\newblock Unreal engine.

\bibitem{ati}
T.~Franke, C.~Attig, and D.~Wessel.
\newblock Assessing affinity for technology interaction – the affinity for technology interaction (ati) scale. scale description – english and german scale version, 07 2017. doi: {{%
10\hspace{.1pt}\discretionary{.}{%
}{.}\hspace{.4pt}13140\discretionary{/}{%
}{/}RG\hspace{.1pt}\discretionary{.}{%
}{.}\hspace{.4pt}2\hspace{.1pt}\discretionary{.}{%
}{.}\hspace{.4pt}2\hspace{.1pt}\discretionary{.}{%
}{.}\hspace{.4pt}28679\hspace{.1pt}\discretionary{.}{%
}{.}\hspace{.4pt}50081}}


\bibitem{grill}
H.~Grillon, F.~Riquier, B.~Herbelin, and D.~Thalmann.
\newblock Virtual reality as a therapeutic tool in the confines of social anxiety disorder treatment.
\newblock {\em International journal on disability and human development}, 5(3):243--250, 2006.

\bibitem{hard}
G.~Hardavella, A.~Gagnat, D.~Xhamalaj, and N.~Saad.
\newblock How to prepare for an interview: Table 1.
\newblock {\em Breathe}, 12:e86--e90, 09 2016. doi: {{%
10\hspace{.1pt}\discretionary{.}{%
}{.}\hspace{.4pt}1183\discretionary{/}{%
}{/}20734735\hspace{.1pt}\discretionary{.}{%
}{.}\hspace{.4pt}013716}}


\bibitem{kell}
R.~Kellems, G.~Yakubova, J.~Morris, A.~Wheatley, and B.~Chen.
\newblock {\em Using Augmented and Virtual Reality to Improve Social, Vocational, and Academic Outcomes of Students With Autism and Other Developmental Disabilities}, pp. 164--182.
\newblock 01 2021. doi: {{%
10\hspace{.1pt}\discretionary{.}{%
}{.}\hspace{.4pt}4018\discretionary{/}{%
}{/}978\discretionary{%
}{-}{-}1\discretionary{%
}{-}{-}7998\discretionary{%
}{-}{-}5043\discretionary{%
}{-}{-}4\hspace{.1pt}\discretionary{.}{%
}{.}\hspace{.4pt}ch008}}


\bibitem{kwon}
J.~Kwon, J.~Powell, and A.~Chalmers.
\newblock How level of realism influences anxiety in virtual reality environments for a job interview.
\newblock {\em International Journal of Human-Computer Studies}, 71:978–987, 10 2013. doi: {{%
10\hspace{.1pt}\discretionary{.}{%
}{.}\hspace{.4pt}1016\discretionary{/}{%
}{/}j\hspace{.1pt}\discretionary{.}{%
}{.}\hspace{.4pt}ijhcs\hspace{.1pt}\discretionary{.}{%
}{.}\hspace{.4pt}2013\hspace{.1pt}\discretionary{.}{%
}{.}\hspace{.4pt}07\hspace{.1pt}\discretionary{.}{%
}{.}\hspace{.4pt}003}}


\bibitem{lerner}
A.~Lerner.
\newblock interviewing.io raises \$3 million to rewrite the rules of engineering hiring with its anonymous platform.
\newblock 2017.

\bibitem{conv}
V.~Logacheva, M.~Burtsev, V.~Malykh, V.~Polulyakh, and A.~Seliverstov.
\newblock {\em ConvAI Dataset of Topic-Oriented Human-to-Chatbot Dialogues}, pp. 47--57.
\newblock 01 2018. doi: {{%
10\hspace{.1pt}\discretionary{.}{%
}{.}\hspace{.4pt}1007\discretionary{/}{%
}{/}978\discretionary{%
}{-}{-}3\discretionary{%
}{-}{-}319\discretionary{%
}{-}{-}94042\discretionary{%
}{-}{-}7\_3}}


\bibitem{ditton}
M.~Lombard, T.~Ditton, L.~Weinstein, and Temple.
\newblock Measuring presence: The temple presence inventory.
\newblock 2009.

\bibitem{looks}
Q.~Min, H.~Sun, X.~Wang, and C.~Zhang.
\newblock How do avatar characteristics affect applicants' interactional justice perceptions in artificial intelligence-based job interviews?
\newblock {\em International Journal of Selection and Assessment}, n/a(n/a). doi: {{%
10\hspace{.1pt}\discretionary{.}{%
}{.}\hspace{.4pt}1111\discretionary{/}{%
}{/}ijsa\hspace{.1pt}\discretionary{.}{%
}{.}\hspace{.4pt}12472}}


\bibitem{imi}
R.~Ryan.
\newblock Control and information in the intrapersonal sphere: An extension of cognitive evaluation theory.
\newblock {\em Journal of Personality and Social Psychology}, 43:450--461, 09 1982. doi: {{%
10\hspace{.1pt}\discretionary{.}{%
}{.}\hspace{.4pt}1037\discretionary{/}{%
}{/}0022\discretionary{%
}{-}{-}3514\hspace{.1pt}\discretionary{.}{%
}{.}\hspace{.4pt}43\hspace{.1pt}\discretionary{.}{%
}{.}\hspace{.4pt}3\hspace{.1pt}\discretionary{.}{%
}{.}\hspace{.4pt}450}}


\bibitem{smith}
M.~Smith, M.~Fleming, M.~Wright, M.~Losh, L.~Humm, D.~Olsen, and M.~Bell.
\newblock Brief report: Vocational outcomes for young adults with autism spectrum disorders at six months after virtual reality job interview training.
\newblock {\em Journal of Autism and Developmental Disorders}, 45, 05 2015. doi: {{%
10\hspace{.1pt}\discretionary{.}{%
}{.}\hspace{.4pt}1007\discretionary{/}{%
}{/}s10803\discretionary{%
}{-}{-}015\discretionary{%
}{-}{-}2470\discretionary{%
}{-}{-}1}}


\bibitem{lulia}
I.-C. Stanica, M.~Dascalu, A.~Moldoveanu, and F.~Moldoveanu.
\newblock Virtual reality training system for improving interview performance.
\newblock pp. 262--267, 04 2018. doi: {{%
10\hspace{.1pt}\discretionary{.}{%
}{.}\hspace{.4pt}12753\discretionary{/}{%
}{/}2066\discretionary{%
}{-}{-}026X\discretionary{%
}{-}{-}18\discretionary{%
}{-}{-}106}}


\bibitem{north}
C.~Ware.
\newblock {\em Information Visualization: Perception for Design}.
\newblock Morgan Kaufmann Publishers Inc., San Francisco, 2\textsuperscript{nd} ed., 2004. doi: {{%
10\hspace{.1pt}\discretionary{.}{%
}{.}\hspace{.4pt}1016\discretionary{/}{%
}{/}B978\discretionary{%
}{-}{-}155860819\discretionary{%
}{-}{-}1\discretionary{/}{%
}{/}50001\discretionary{%
}{-}{-}7}}


\bibitem{ar-au}
Q.~Xu, S.-c.~S. Cheung, and N.~Soares.
\newblock Littlehelper: An augmented reality glass application to assist individuals with autism in job interview.
\newblock In {\em 2015 Asia-Pacific Signal and Information Processing Association Annual Summit and Conference (APSIPA)}, pp. 1276--1279. IEEE, 2015.

\end{thebibliography}
\end{document}